\documentclass[
]{ceurart}

\usepackage{listings}
\usepackage{minted}
\usepackage[show]{chato-notes}
\usepackage{amsmath}
\usepackage{subcaption}

\usepackage{listings}
\usepackage{xcolor}
\usepackage{fontawesome5}
\usepackage{amsthm}

\usepackage{stmaryrd}

\newcommand{\lBrack}{\llbracket}
\newcommand{\rBrack}{\rrbracket}

\definecolor{codeblue}{HTML}{1E4F8C}    %
\definecolor{codepurple}{HTML}{6A1B9A}  %
\definecolor{codegreen}{HTML}{1B5E20}   %
\definecolor{codeorange}{HTML}{A84300}  %
\definecolor{codegray}{HTML}{555555}    %
\definecolor{codefg}{HTML}{111111}

\lstdefinestyle{vscodedark}{
    backgroundcolor=\color{lightgray!20},
    basicstyle=\ttfamily\small\color{codefg},
    keywords={pt,Experiment, dataset,get_topics,get_qrels, from, import},
    keywordstyle=\color{codeblue},
    keywords=[2]{BM25,monoT5,duoT5},
    keywordstyle=[2]\color{codepurple},
    keywordstyle=[3]\color{codegreen},
    stringstyle=\color{codeorange},
    commentstyle=\color{codegreen},
    numberstyle=\tiny\color{codegray},
    breaklines=true,
    numbers=left,
    frame=single,
    showstringspaces=false
}

\begin{document}

\copyrightyear{2026}
\copyrightclause{Copyright for this paper by its authors.
  Use permitted under Creative Commons License Attribution 4.0
  International (CC BY 4.0).}

\conference{ReNeuIR'26: Workshop on Reaching Efficiency in Neural Information Retrieval,
  July 24, 2026, Melbourne, Australia}

\title{Trie-based Experiment Plans for Efficient IR Pipeline Experiments}

\author{Irene Anu}[%
orcid=0009-0003-0157-3812,
email=2772146I@student.gla.ac.uk,
]
\address{University of Glasgow, United Kingdom}

\author{Craig Macdonald}[%
orcid=0000-0003-3143-279X,
email=craig.macdonald@glasgow.ac.uk,
]

\begin{abstract}
Search engines are often formulated as cascading pipelines, where successive stages combine the results of different retrievers, and iteratively refine the ranking of candidate documents to obtain a final ranking, which can be presented to a user, or provided as context to an LLM. Such pipelines can be complex to evaluate in an end-to-end manner, necessitating measurement of Recall of early stages, and Precision of later stages, which are often interchangeable. PyTerrier is ideal for building and evaluating cascading retrieval pipelines, due to its declarative nature for pipeline construction and wide ecosystem of retrievers and rerankers. However, comparative evaluation of pipelines can be expensive due to repeated components. In this work, we describe the use of a trie data structure to formulate an {\em experiment plan} for comparative pipeline experiments that enhances experiment efficiency compared to a sequential ``linear" plan. Empirically, on a demonstration experiment involving BM25, MonoT5 and DuoT5 on MSMARCO v2, we observe a 26\% reduction in experiment duration. Finally, we report on a user study of undergraduate and postgraduate research students' use of the experiment plans. 
\end{abstract}

\begin{keywords}
Information Retrieval Experiments
\end{keywords}

\maketitle

\section{Introduction}

The scale of corpora available to search is constantly growing which necessitates that modern Information Retrieval (IR) systems balance the accuracy of the retrieved results with the computational cost to produce them~\citep{tonellotto2013efficient}. Retrieval systems are often formulated as pipelines, notably {\em cascading} ranking pipelines that progressively re-rank a candidate set of documents with more expensive retrieval models \citep{DBLP:journals/ftir/Liu09,wang2011cascading,macavaney2022adaptive}. In such pipelines, a dense retriever is used to obtain an initial candidate set which are further refined by neural rerankers such as mono, duo~\cite{DBLP:journals/corr/abs-2101-05667} or even listwise rerankers~\cite{pradeep2023rankvicunazeroshotlistwisedocument}.

With the prevalence of such pipeline architectures, PyTerrier allows composition of pipelines in a declarative manner with a suite of function objects collectively called \textit{transformers}. In this Python formulation, operators are used to indicate combinations of Transformers -- for instance, {\tt $\gg$} and {\tt $\%$} are used to indicate composition of Transformers and rank cutoff operations respectively. This allows the rapid formulation of pipeline variants with alternative retrieval backends, ranking or reranking models, query reformulators, etc. - indeed there exists a wide ecosystem of plugins addressing functionality including dense retrieval~\cite{macdonald2026replicability,10.1145/3459637.3482013}, rerankers, and RAG~\cite{macdonald2025constructing}. Such pipelines can also be easily visualised~\cite{DBLP:conf/ecir/LionisMM26} (see Figure~\ref{fig:schematic}), or evaluated using test collections obtained from ir\_datasets~\cite{irdatasets} (a lightweight library that provides easy access to a variety of IR corpora and query sets).

Indeed, it is common for researchers to construct and compare multiple pipeline variants to evaluate the effectiveness of new techniques. However, a practical challenge that arises is the need to re-run large shared components of these pipelines, resulting in substantial redundant computation. For example,~\citet{DBLP:conf/ecir/PenhaCH22} conducts multiple experiments that all rely on the same first-stage retriever BM25, leading to repeated execution of identical retrieval steps. This inefficiency emphasises the need for more efficient IR experimentation methods. 

\looseness -1 In this work, we propose the notion of {\em experiment plans} -- the decomposition of a set of pipeline comparisons involving shared components into a minimal set of executions. For instance, a {\em linear} plan would execute each pipeline sequentially. Earlier work~\cite{macavaney2025precomputation} identified and reused longest common prefixes across all pipelines in the experiment to avoid repeated execution of identical initial stages. Going further, in this work, we propose the formulation of a {\em tree} experiment plan, that identifies and reuses all possible shared prefixes, created by instantiating a trie data structure to detect overlapping pipeline prefixes. Overall, this work contributes: (1) formulation of optimal experiment plans for pipeline retrieval experiments using radix trees; (2) demonstration of the efficiency benefits for realistic experiments upon the MSMARCO test collection; and (3) summary of a user study upon using the changes.

The structure of the remainder of the paper is as follows: Section~\ref{sec:related} discusses related work; Section~\ref{sec:pyterrier} provides an introduction to the PyTerrier data model, operators and declarative experimentation; Section~\ref{sec:method} introduces our experimental plan; Section~\ref{sec:exp} demonstrates the benefit of the plans when conducting experiments; Section~\ref{sec:users}  provides an overview of feedback from a user study; Section~\ref{sec:conc} provides concluding remarks.

\section{Related Work}\label{sec:related}

Much of the IR literature focuses on improving the efficiency (and/or effectiveness) of executing individual queries~\cite{fntir2018efficient}, often through techniques such as cascading retrieval pipelines~\cite{DBLP:journals/ftir/Liu09,wang2011cascading}. In contrast, efficient execution of inference experiments receive much less attention, despite the substantial redundancy that arises when evaluating many related pipeline variants. The most relevant work in PyTerrier is the current optimisation (described in ~\cite{macavaney2025precomputation}) that identifies the Longest Common Prefix (LCP) of stages that are common across all pipelines in an experiment. This approach can reduce experiment duration when every pipeline begins with the same sequence of initial transformers, however it brings no benefit when components are shared across subsets of pipeline rather than the entire set. 

Database systems offer a useful perspective for addressing this gap in identifying common subsequences across query constructs and ensure they are executed only once. Multi-query optimisation in database systems detects common sub-expressions by examining the logical query expressions, allowing these shared parts to be evaluated once and reused across all queries that access the same data~\cite{sellis1988multiple}. This optimisation provides a useful conceptual model for reducing redundant computation across pipelines that share common prefixes, but its direct application to PyTerrier is constrained by operator semantics. The semantics of an IR pipeline are determined by its leftmost components, hence only prefix-level reuse is safe, once pipelines diverge into different transformations, their computations can no longer be shared without altering behaviour~\cite{macavaney2025precomputation}.

In addition to work on multi-query optimisation, database systems introduce logical query plans as a coherent framework for expressing how a query is organised internally.  Database systems translate a user query through several compilation stages ultimately producing an optimised logical query plan~\cite{382293}. This abstraction separates the logical structure of the computation from the details of how it is executed, enabling optimisation and reuse of shared operations. PyTerrier similarly processes queries through a sequence of components such as retrievers, rerankers, feature transformers and fusion operators, yielding a broadly comparable staged architecture. We take forward this offline plan abstraction to IR pipelines. Instead of algebraic optimisation, our approach focuses on providing an alternative tree execution plan to the existing execution strategy, that executes the same experiments more efficiently using radix trees. 

Beyond plan-based approaches, prior work also explores how database queries can be represented structurally. One common approach is to formulate the execution plan as a directed acyclic graph (DAG)~\cite{Neumann2005a}. In dataflow systems such as Apache Spark~\cite{sparksql}, the execution of a job is organised as a directed acyclic graph, where each vertex corresponds to a stage and edges capture the flow of data between stages~\cite{9894699}. However, Spark is a generic dataflow architecture and our experience using Terrier-Spark~\cite{sparkterrier} indicate it is inefficient when applied to information retrieval querying workflows. Although this limitation exists, its use of DAGs illustrates how representing a workflow as a graph can make structural relationships explicit. More relevant, QPipe~\cite{harizopoulos2005qpipe} introduces simultaneous pipelining, in which a relational operator that produces identical tuples for multiple concurrent queries is executed once and its output is broadcast to all consuming operators. Our work is conceptually aligned with this idea, where we also exploit shared computation across multiple query pipelines. However, instead of sharing relational operators at runtime, we identify prefix overlaps across IR pipelines and execute these shared stages once.

Within IR, we acknowledge the TIREx platform~\cite{FrobeRMDRB0HP23} -- which builds upon PyTerrier interfaces -- and allows different stages of a retrieval pipeline to be dockerised for reproducibility, and their results for a wide variety of test collections archived. TIREx enables intermediate results to be cached and shared across stages. This emphasis on structural decomposition and reusable stages is consistent with our objective of exploiting structural overlap between pipelines, enabling reuse of intermediate results and reducing redundant work. However, our focus is upon fast adhoc experimentation local to the researcher, rather than the archived preservation of results that TIREx offers.

Despite the breadth of prior research, none of these approaches are directly applicable to identify common prefix overlaps in IR pipelines but they provide valuable insights that collectively motivate the design of our proposed approach.

\section{Overview of PyTerrier}\label{sec:pyterrier}
PyTerrier represents indexing and retrieval components such as rankers, rerankers and feature extractors as {\em transformers}, each of which takes a dataframe as input and produces a transformed dataframe as output. Transformations typically act upon known standard dataframe types, such as $\mathcal{D}$ documents, $\mathcal{Q}$ queries, $\mathcal{R} \subset \mathcal{D} \times  \mathcal{Q}$ retrieved documents, and $\mathcal{A}$ question answers, etc. For instance, a retrieval transformer $t$ may be executed upon a set of queries $Q \subset  \mathcal{Q}$:
$$
\lBrack t \rBrack (Q)
$$
where $\lBrack \cdot \rBrack(\cdot)$ denotes invocation of the bracket expression with the parenthesised parameter(s).

To combine different transformers, PyTerrier implements a number of operators defined on transformers, that allow them to be expressed in a declarative manner~\cite{macavaney2025precomputation}. For instance, the $\gg$ operator -- known as ``then'' or ``compose'' -- is used to allow multi-stage pipelines through transformer function composition, and is defined as:
\begin{align}
\lBrack t_1 \texttt{>>} t_2 \rBrack(Q) := \lBrack t_2 \rBrack ( \lBrack t_1\rBrack(Q)) \nonumber
\end{align}
and implemented in Python for PyTerrier transformer objects, by use of operator overloading. Other operators exist within PyTerrier, including rank cutoffs ($\%$) and linear combination ($+$), which can be applied to transformers within the pipeline~\cite{10.1145/3459637.3482013}. 

\begin{figure}
    \centering
    \includegraphics[width=0.85\linewidth]{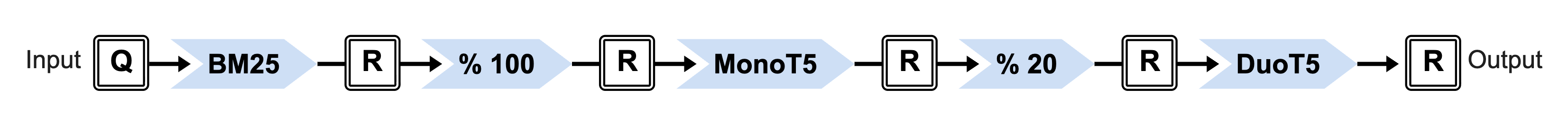}
    \caption{PyTerrier generated {\em schematic}~\cite{DBLP:conf/ecir/LionisMM26} visualisation of the pipeline from Listing~\ref{lst:single_pipeline}.}
    \label{fig:schematic}
\end{figure}

This notation can be used to express complex retrieval pipelines concisely in a declarative manner demonstrating their underlying conceptual behaviour \citep{10.1145/3459637.3482013}. The example shown in Listing~\ref{lst:single_pipeline} employs a multi-stage retrieval pipeline in PyTerrier. The initial stage retrieves the top-100 candidate documents using BM25, applies MonoT5 to re-rank this candidate set and finally applies DuoT5 pairwise re-ranking to the top-20 results from the candidate set, to compute refined relevance scores and produce the final ranking. Pipelines can be visualised as {\em schematics}~\cite{DBLP:conf/ecir/LionisMM26}, allowing the stages and data type transformation to be easily inspected, as shown in Figure~\ref{fig:schematic}.

\begin{lstlisting}[float,caption={Multi-stage pipeline in PyTerrier}, label={lst:single_pipeline}, language=Python, style=vscodedark]
from pyterrier_t5 import MonoT5ReRanker, DuoT5ReRanker
BM25 = index.bm25(include_fields=['text'])
monoT5 = MonoT5ReRanker() # loads castorini/monot5-base-msmarco
duoT5 = DuoT5ReRanker() # loads castorini/duot5-base-msmarco
pipeline = BM25 % 100 >> monoT5 % 20 >> duoT5
\end{lstlisting}

\looseness -1 PyTerrier extends its flexible pipeline construction model with a declarative API for evaluating and comparing retrieval pipelines. This evaluation interface allows researchers to compare different retrieval pipelines across multiple query sets and effectiveness measures without manually managing intermediate files or relying on evaluation tools such as \emph{trec\_eval}, as all intermediate results are handled internally~\citep{macdonald2025constructing}. Central to this framework is the \texttt{pt.Experiment()}  function, which provides an abstraction for comparative evaluation. It accepts four key inputs: a list of retrieval pipelines to be compared, the set of queries on which they should be evaluated, the corresponding relevance assessments, and the evaluation measures to compute. It outputs a dataframe containing the computed measures for each system, providing basis for comparison and further analysis~\citep{macavaney2025precomputation}. An example usage of \texttt{pt.Experiment()} can be seen in Listing~\ref{lst:ex_pipeline}, illustrating how multiple retrieval pipelines can be evaluated within a unified experimental setup. These retrieval pipelines correspond to the example pipeline introduced in Listing~\ref{lst:single_pipeline}, extending the shared {\tt BM25 \% 100} stage with deep neural re-ranking stages.

\begin{lstlisting}[float, caption={Example experiment in PyTerrier. \% denotes rank cutoff.}, label={lst:ex_pipeline}]
pt.Experiment(
    [
     BM25 % 100, 
     BM25 % 100 >> monoT5, 
     BM25 % 100 >> monoT5 % 10 >> duoT5, 
     BM25 % 100 >> monoT5 % 20 >> duoT5
    ],
    dataset.get_topics(), 
    dataset.get_qrels(),
    [NDCG@10])
\end{lstlisting}

By integrating dataset handling, pipeline execution and evaluation within a single interface, PyTerrier removes much of the operational complexity traditionally associated with IR experimentation, such as, moving results from different stages, writing results to files before invoking {\tt trec\_eval}, significance testing, batching of queries etc.%

However, a key disadvantage of a declarative approach is the duplication of execution -- for instance, in the example experiment above BM25 would be executed four times for each query; recent work~\cite{macavaney2025precomputation} aimed to avoid such duplication. This method applies the longest common prefix algorithm to detect the single longest pipeline shared by all pipelines within the experiment, which can be precomputed and reused by all pipelines. This optimisation, however, would not avoid re-execution of the MonoT5 stage in Listing~\ref{lst:ex_pipeline} as it is not a common prefix for the pipeline {\tt BM25 \% 100}. In this work, we address such a limitation by developing more generic experiment plans that avoid redundant recomputation.

\section{Experiment Planning for IR Pipelines in PyTerrier}\label{sec:method}

This paper proposes the notion of {\em experiment plans}, that aim to define an efficient traversal of all transformers within an experiment, reducing duplicate execution efforts and thereby increasing experimental efficiency. We formalise the experimental planning problem as a traversal of a directed acyclic graph in Section~\ref{ssec:method:formalism}; motivate our instantiation of a tree experiment plan using a radix tree in Section~\ref{ssec:method:tries}; describe our implementation in PyTerrier and the advantages it brings in Section~\ref{ssec:method:impl}; finally, Section~\ref{ssec:method:discussion} concludes with a broader discussion of the contribution, highlighting the benefits and novelty of our proposed approach.

\subsection{Definition}\label{ssec:method:formalism}

Let $\mathcal{P}$ \ be a finite set of transformer pipelines.  
Each pipeline $P_i \in \mathcal{P}$ \ is an ordered sequence of stages
\[
P_{i,1} \gg P_{i,2} \gg \dots \gg P_{i,\lVert P_i \rVert},
\]
where $\lVert P_i \rVert$ denotes the number of stages of the pipeline and $P_{i,l}$ its $l$-th stage.

For $0 \le l \le \lVert P_i \rVert$, let {\em pref} $(P_i, l)$ denote the prefix consisting of the first $l$ stages of $P_i$.  
Two pipelines $P_i$ and $P_j$ share a prefix of length $l$ if {\em pref} $(P_i, l)=$ {\em pref} $(P_j, l)$. For example, if $\mathcal{P}\phantom{x}$ is defined from Listing~\ref{lst:ex_pipeline}, we have {\em pref} $(P_i, 2) =$ {\tt BM25 \% 100} for every pipeline $P_i \in \mathcal{P}$, since the first two stages are identical across all four pipelines. Moreover, for the latter three pipelines, {\em pref} $(P_i, 3) =$ {\tt BM25 \% 100 >> monoT5}, showing that they share a longer common prefix. Such comparison and identification of common prefixes is possible because PyTerrier allows to define transformer equality, allowing to test whether two pipeline stages are identical, as noted in~\cite{macavaney2025precomputation}.

We define the global execution directed acyclic graph (DAG) of $\mathcal{P}$ \ as $G=(V,E)$.  
Each node $v \in V$ corresponds to a stage in some pipeline and is labelled by $\lambda(v)=P_{i,l}$. 
The unique path from the root to $v$ is $\rho(v) =$ {\em pref} $(P_i, l)$, the finite sequence of stages up to $P_{i,l}$.
An edge $(u,v) \in E$ is present iff
\[
\rho(v) = \rho(u) \gg \lambda(v),
\]
i.e., an edge between nodes $u$ and $v$ exists if the path to $v$ is the path to $u$ extended by $v$. 
Thus, if $\rho(u)=pref(P_i,l)$ and $\rho(v)=pref(P_i,l+1)$, the edge $(u,v)$ represents the execution step in which the output of $u$ becomes the input to $v$, corresponding to applying the next stage of the pipeline.

Our DAG allows us to identify pipelines that share common prefixes by reusing nodes -- compute the results for those nodes once and passing them onto multiple child nodes.
A node corresponding to $P_{i,l}$ is reused if and only if there exists a node $n$ such that $\rho(n) = pref(P_i,l)$. %
Nodes are shared across pipelines whenever the corresponding stages are reached through identical pipeline prefixes. Consequently, a common prefix occurring in multiple pipelines is represented by a single node in the execution DAG.

\begin{figure}[tb]
    \centering

    \begin{subfigure}{0.45\textwidth}
        \centering
        \includegraphics[width=\linewidth]{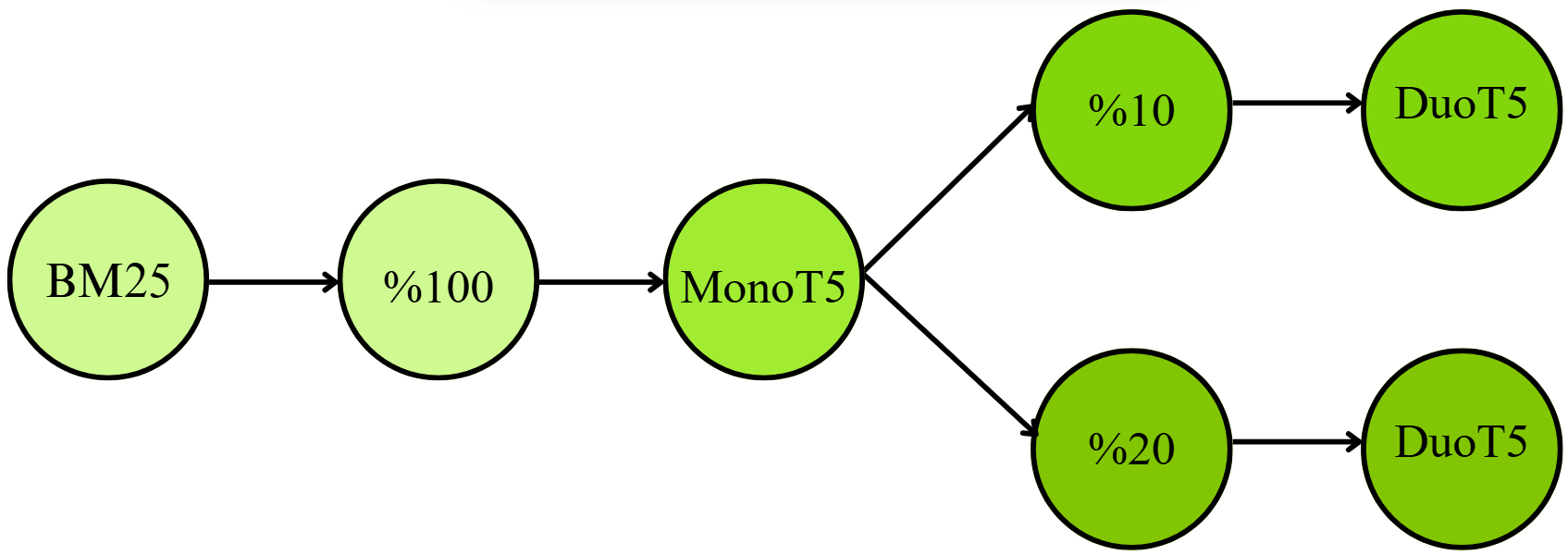}
        \caption{Execution DAG diagram.}
        \label{fig:dag}
    \end{subfigure}\hspace{10mm}
    \begin{subfigure}{0.45\textwidth}
        \centering
        \includegraphics[width=\linewidth]{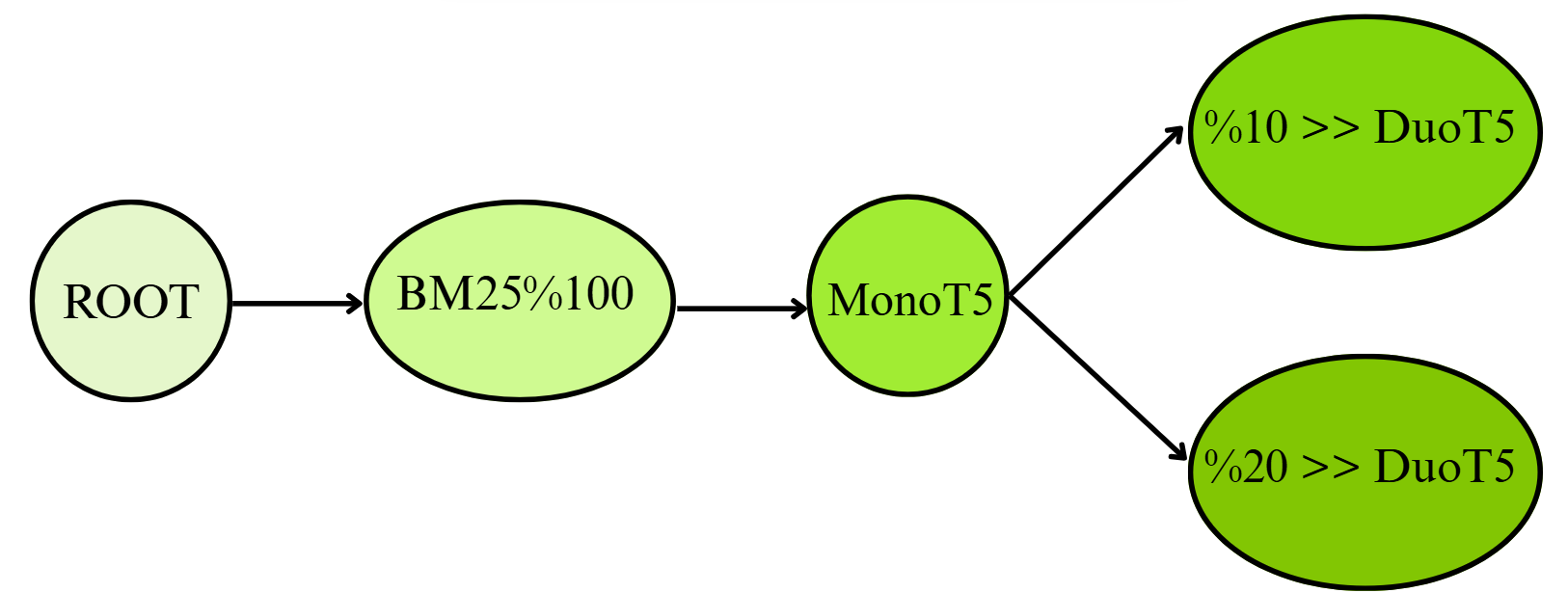}
        \caption{radix tree instantiation of DAG in Figure~\ref{fig:dag}.}
        \label{fig:radix}
    \end{subfigure}

    \caption{Structural comparison between execution DAG and radix tree implementation derived from the pipelines in Listing~\ref{lst:ex_pipeline}.}
    \label{fig:dag_trie}
\end{figure}

The execution DAG for the pipeline set in Listing~\ref{lst:ex_pipeline} is illustrated in Figure~\ref{fig:dag}. Nodes are reused whenever their corresponding prefixes are identical, such that common prefixes {\tt BM25 \% 100} and {\tt BM25 \% 100 >> monoT5} appear only once in the graph despite being shared by multiple pipelines. Consequently, the shared computations represented by the common prefixes %
are executed once and their intermediate results are reused by all child nodes, eliminating redundant execution while preserving the output of each pipeline.

A global execution DAG is correct if the execution of shared prefixes does not alter the output of any pipeline. In particular, for every pipeline $P_i \in \mathcal{P}$, the result obtained by evaluating the corresponding path in the global execution DAG should be identical to the result that would be obtained by executing that pipeline linearly in sequential order, i.e. 
$$
\forall i, \ \pi_i(Exec(G,Q)) =  \lBrack Pi \rBrack(Q)
$$
where $\pi_i(\cdot)$ denotes the projection of the results for the $i$-th pipeline, and $Exec(\cdot)$ denotes full traversal of execution graph $G$.

Shared execution affects only the execution order and not the output of any pipeline. Each pipeline is evaluated along its corresponding root-to-leaf path in the execution DAG, preserving the original ordering of transformers (retrieval, reranker, answer generation etc.). Nodes are identified by their execution paths $\rho(v)$, each distinct intermediate computation state is represented by a unique node in the graph. A valid execution corresponds to a topological traversal of the DAG in which each node is executed exactly once after all its predecessors have been executed. The resulting intermediate outputs are then reused by all outgoing edges, ensuring that shared prefixes are not recomputed while preserving pipeline semantics.

\looseness-1 The challenge then becomes how to build such a DAG and traverse it. The keen observer will note that due to the prefix property, the execution graph always forms a prefix tree. In the next section, we show how a prefix tree or {\em trie} data structure can be used to construct and execute the tree-structured graph.

\subsection{Tries as Execution Plans}\label{ssec:method:tries}

Pipelines in PyTerrier exhibit sequence‑like behaviour because they expose a container interface that supports iteration and indexing which support the inspection of individual pipeline stages. These operations allow composite pipelines to be treated in the same way as strings, with each transformation stage corresponding to an element in an ordered sequence. This behaviour provides us a natural motivation to investigate whether pipelines can be analysed using data structures and algorithms typically applied to strings, most notably tries.

\looseness -1 A trie, also known as a prefix tree, is a structured graph, in that each node has a unique parent and that merging occurs only when two sequences share an identical prefix~\cite{larsson1999structures}. As a pipeline can be treated as a finite sequence of transformers, a  trie structure allows to identify the overlapping prefixes among pipelines.
The resulting graph compactly represents the entire pipeline set while preserving the execution order of every individual pipeline. Traversal of the trie, i.e.\ executing each node and passing its output as input to each child node, is equivalent to a complete execution of the full set of pipelines $\mathcal{P}$. Together, these properties allow the trie to function as a global execution DAG. 

\looseness-1 The different types of tries we considered were the prefix tree~\cite{Karimov2020}, suffix trie \cite{gunasinghe2013adaptive} and compressed (radix) trie \cite{Morrison68}. Among these, the radix tree is an ordered, space-optimised trie structure. Any node with only a single child is merged with that child producing edges with shared transformers composed together, unless it is a terminal node, in which case it is preserved to ensure that pipelines that are strict prefixes of other pipelines remain explicitly represented. Compared to a normal trie, this reduces the height and number of nodes while preserving the structural properties of the underlying trie. For pipeline execution, a radix tree  provides a practical advantage, in that the compressed nodes are not decomposed into different stages, and as a result, the intermediate outputs for those stages do not need to be stored. For example, in Figure~\ref{fig:dag_trie} when a common prefix such as {\tt BM25 \% 100} is represented as a single compressed edge, only the top‑100 results need to be stored in the memory while the child nodes are executing. For a conventional trie, the complete output of BM25 would be retained. Thus, using a radix tree avoids retaining intermediate results that would never be reused and reduces the storage overhead of the underlying radix tree structure. This motivates our choice of a radix tree representation as the core abstraction that enables offline experiment planning, shared prefix execution and the elimination of redundant computation.

Figure~\ref{fig:dag_trie} highlights the correspondence between the formal DAG defined in Section~\ref{ssec:method:formalism} and its radix‑trie instantiation. Specifically, Figure~\ref{fig:radix} shows the radix tree representation of the pipelines in Listing~\ref{lst:ex_pipeline} and shared prefixes are merged into single nodes, with branching occurring only at points of divergence. Indeed, as we show later in Section~\ref{sec:exp}, for an experiment of the form shown in Listing~\ref{lst:ex_pipeline}, repeated execution of  prefix components like BM25 and MonoT5 are eliminated, to markedly benefit experiment execution time.

\subsection{Implementation in PyTerrier}\label{ssec:method:impl}

To instantiate the formal definition of the prefix DAG, we implement the DAG as a radix tree, which forms the basis of an improved experiment planning, building upon the earlier implementation that used longest common prefix~\cite{macavaney2025precomputation} for reducing extraneous computation. Each node in the trie corresponds to a unique pipeline prefix and stores the ordered sequence required to reach that point. Nodes additionally record whether they correspond to the terminal stage of a pipeline. Terminal nodes mark the complete execution of a particular pipeline, at which point evaluation measure computation is invoked over its output. We also note that terminal nodes are not necessarily leaf nodes. A pipeline may terminate at a node even if other pipelines extend beyond it. In such case, the node is terminal because it represents a complete pipeline, but it is not a leaf node because it still has children corresponding to longer pipelines that share its prefix.

The radix tree execution strategy is exposed in PyTerrier as the tree execution plan, enabled by setting the parameter {\tt plan='tree'} in \texttt{pt.Experiment()}. The tree execution plan begins by inserting all pipelines into the radix tree structure, ensuring that shared prefixes are represented once. Then a preorder traversal is performed of the resulting structure, such that each node is executed exactly once. This yields a minimal execution plan that avoids redundant computation while preserving the semantics of each pipeline.\footnote{This mode may be enabled by default in future versions of PyTerrier.} Listing~\ref{lst:pipeline} shows a concrete example of how this plan is applied in PyTerrier.

During traversal, the interim results  for a given non-terminal node must be retained in memory until all children nodes evaluated. A set of 1000 results for a single query requires a dataframe of around $\sim$200KB. Trees with many different branches, or with larger number of queries and/or large numbers of results will carry larger overheads.\footnote{Initial experiments on the example pipelines used in this paper detected no measurable memory overhead.} 
The number of queries in an evaluation batch can easily be controlled in a {\tt pt.Experiment()} invocation should this be a problem. Traversal also records the execution time of each node, in order to provide estimates of mean response time.

\begin{lstlisting}[float, caption={Tree experiment plan integration in PyTerrier. {\tt plan='tree'} enables the trie-based plan. In a notebook environment, the visualisation shown in Figure~\ref{fig:pipeline_progress} is automatically displayed.}, label={lst:pipeline}]
pt.Experiment(
    [
     BM25 % 100, 
     BM25 % 100 >> monoT5, 
     BM25 % 100 >> monoT5 % 10 >> duoT5, 
     BM25 % 100 >> monoT5 % 20 >> duoT5
    ],
    dataset.get_topics(), 
    dataset.get_qrels(),
    [NDCG@10], 
    plan = 'tree')
\end{lstlisting}

\begin{figure}[tb]
    \centering

    \begin{subfigure}{0.9\textwidth}
        \centering
        \includegraphics[width=\linewidth]{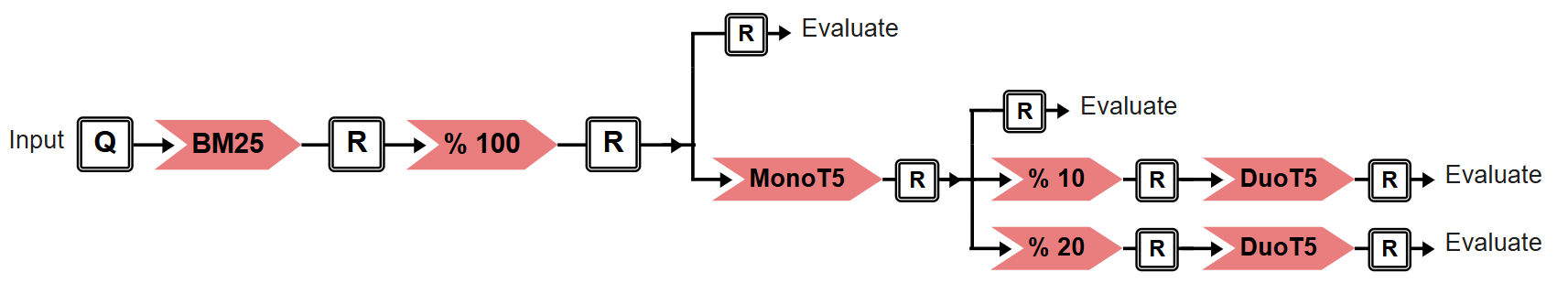}
        \caption{Before execution: all transformers are red.}
        \label{fig:pipe1}
    \end{subfigure}

    \vspace{0.5em}

    \begin{subfigure}{0.9\textwidth}
        \centering
        \includegraphics[width=\linewidth]{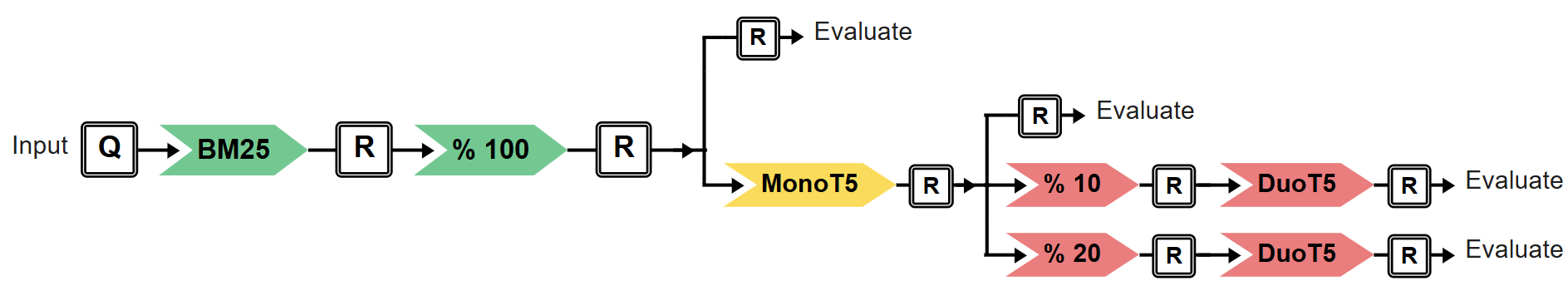}
        \caption{During execution: completed transformers in green, those currently being executed are in yellow.}
        \label{fig:pipe2}
    \end{subfigure}

    \vspace{0.5em}

    \begin{subfigure}{0.9\textwidth}
        \centering
        \includegraphics[width=\linewidth]{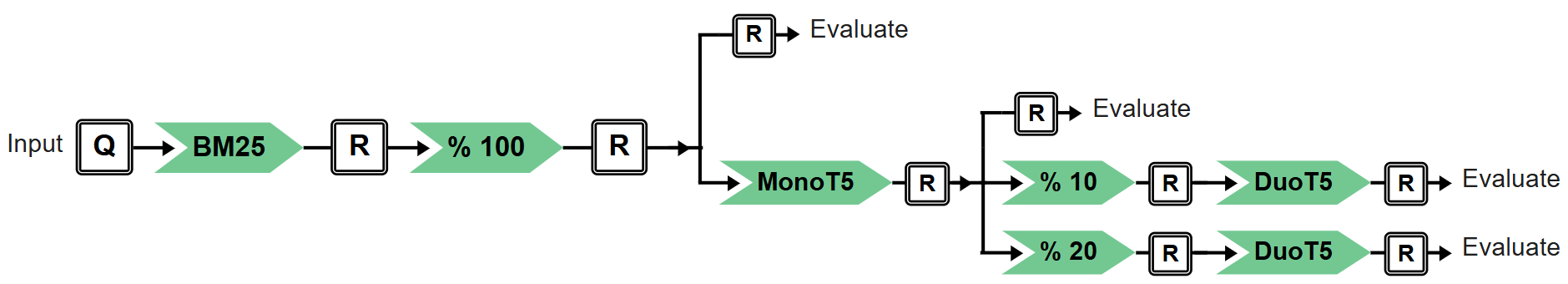}
        \caption{After execution: all transformers are green.}
        \label{fig:pipe3}
    \end{subfigure}

    \caption{Visualisation of experiment execution progress. This visualisation is shown in notebook environments whenever {\tt plan='tree'} is set.}
    \label{fig:pipeline_progress}
\end{figure}

Finally, building upon PyTerrier's recent schematic visualisation capabilities~\cite{DBLP:conf/ecir/LionisMM26}, the execution of the radix tree is visualised automatically.
Indeed, as the experiment progresses with execution of the nodes of the trie, the visualisation highlights which prefixes have been executed, which remains pending, and how the pipelines branch and diverge. As the HTML visualisation renders each stage in place, Figures~\ref{fig:pipe1}, \ref{fig:pipe2} and \ref{fig:pipe3} show snapshots of the same experiment defined in Listing~\ref{lst:pipeline} at three different stages: before execution, during partial execution and after all pipelines have been executed. The visualisation is stage‑oriented, in that it displays each pipeline component as an individual node, reflecting the conceptual structure of the pipeline rather than the compressed representation of the radix tree. All of the trie-based experiment plan functionality has recently been integrated into PyTerrier.

\subsection{Discussion}\label{ssec:method:discussion}
The radix tree experiment plan has observable benefits in comparison to the current linear execution plan. In the current implementation PyTerrier uses longest common prefix (LCP)~\cite{macavaney2025precomputation}, where only the prefix shared by all of the pipelines is identified. Once the first divergence occurs, the optimisation stops, even if subsets of pipelines share additional prefixes deeper in their structure. A trie is instead able to identify every shared prefix, not just the global one. Whenever a set of pipelines share an initial subsequence, the radix trie represents it as a single node. This ensures that reuse is discovered at all branching points. The radix trie preserves the full hierarchy of shared and divergent prefixes giving a more accurate structure for the computation that can be reused. The radix trie-based experiment plan captures all valid opportunities for reuse, preserves the full prefix structure of the pipelines, and provides a principled foundation for efficient execution.

Readers may consider the possibility of reuse of results by reranker stages - this is inherently possible for pointwise rankers like MonoT5, and made accessible through transformer caching~\cite{macavaney2025precomputation} - i.e. under the {\em explicit} control for researcher. On the other hand, pairwise and listwise rerankers are functionally dependent on all input documents for a given query, and hence are less amenable to caching under the transformer interface.

We believe this is the first work to completely address the complex dependencies between different retrieval workflows automatically, as this challenge only appears during offline experiments where multiple pipeline variants need to be executed and evaluated. The resulting implementation has addressed the low efficiency disadvantage of declarative experimental workflows. 

Next, in Section~\ref{sec:exp} we report the results of the empirical experiments conducted to demonstrate the benefit of our tree-based experiment plan to researchers. It further improves users' understanding of the underlying execution process through the visualisation of experiment progress. Later, in Section~\ref{sec:users}, we report on the experiences of both novice and expert PyTerrier users.

\section{Experiments}\label{sec:exp}
We now present an experiment we conducted to demonstrate the efficiency benefits of our proposed tree experiment plan. The experiment compares all three execution strategies (linear, linear with LCP precomputation~\cite{macavaney2025precomputation}, and tree experiment plans) for evaluating variants of complex IR pipelines in PyTerrier. In particular, we conduct evaluation of the example pipelines defined in Listing~\ref{lst:ex_pipeline}.

\looseness -1 All experiments are conducted using an Intel Xeon Gold 6244 CPU @ 3.60GHz and a Nvidia Titan RTX. BM25 retrieval is conducted using the Terrier retrieval backend, while we use the {\tt pyterrier\_t5} library for MonoT5 and DuoT5. Our notebook is available at \faGithubSquare{}~\url{https://github.com/terrier-org/pyterrier/blob/master/examples/experiments/trie-execution-demo-msmarco-v1v2.ipynb}. 

\looseness -1 We report the time taken to conduct the same experiment on MSMARCO v1~\cite{nguyen2016ms,craswell2019overview} and v2~\cite{craswell2025overviewtrec2021deep} corpora, using the TREC 2019 and TREC 2021 Deep Learning Track query sets, respectively. We report the overall experiment duration, however, we also verify that the reported effectiveness (nDCG@10) is identical across the three experiment plans. Note that the tree-based plan has no hyperparameters to tune.

Experiment durations for MSMARCO v1 and v2 are reported in Table~\ref{tab:experiments}, as well as percentages wrt.\ the default unoptimised linear baseline plan.
On examining the results for MSMARCO v1, we observe that the linear plan with prefix precomputation achieves 1\% decrease relative to the baseline. This represents only a marginal improvement, indicating that although the {\tt BM25 \% 100} prefix is reused, the more expensive neural reranking stages, MonoT5 and DuoT5, still execute independently for each pipeline, leaving the overall execution structure largely unchanged. In contrast, the tree-based plan achieved a 19\%  decrease in execution time relative to the baseline, by eliminating redundant computation across all shared prefixes. This constitutes a substantial reduction in total execution time and demonstrates that the tree-based experiment plan is more efficient at exploiting shared prefixes across all pipelines, preventing re-execution of shared stages.

A comparable pattern can be seen in MSMARCO v2, where  the linear plan with LCP lowers the runtime to 94\% of linear baseline plan, whereas the tree‑based plan achieves a considerably lower relative execution time of 74\%, corresponding to a 26\% decrease relative to the baseline. The higher relative reduction observed for MSMARCO v2 is due to the fact that the baseline linear plan performs repeated BM25 retrieval over a substantially larger index than v1, making redundant retrieval disproportionately expensive. In the tree-based experiment plan, this retrieval is executed only once as a shared prefix, while the reduced recomputation of the neural stages further contributes to the overall efficiency gains.

Overall, the results show that the tree-based plan consistently outperforms both linear execution strategies by avoiding unnecessary recomputation of overlapping neural stages. Notably, across all experiments, the nDCG@10 evaluation measurements produced by the tree-based strategy were identical to those of the linear and LCP baselines, confirming that our proposed tree-based experiment plan preserves effectiveness while improving efficiency. Indeed, on MSMARCO v2 experiments, we observe a 26\% decrease in experiment duration. However, these benefits will clearly vary from experiment to experiment, depending on the nature of the shared components.
\begin{table}[tb]
    \centering
    \begin{tabular}{c|c|cc|cc}
        \toprule
        & & \multicolumn{2}{c|}{MSMARCO v1} & \multicolumn{2}{c}{MSMARCO v2} \\
        \# & Experiment Plan & Execution Time & $\Delta$ & Execution Time & $\Delta$ \\
        \midrule
        (1) & Linear without LCP & 27min 27s & -- & 30min 18s & -- \\
        (2) & Linear with LCP    & 27min 11s & 99\% & 28min 23s & 94\% \\
        (3) & Tree               & 22min 15s & 81\% & 22min 29s & 74\% \\
        \bottomrule
    \end{tabular}
    \caption{Execution times for an experiment comparing \texttt{[BM25 \% 100, BM25 \% 100 >> monoT5, BM25 \% 100 >> monoT5 \% 10 >> duoT5, BM25 \% 100 >> monoT5 \% 20 >> duoT5]} on queries from the TREC 2019 \& 2021 Deep Learning tracks. Relative experiment execution time compared to row (1) are denoted by $\Delta$.} 
    \label{tab:experiments}
\end{table}

\section{User Study}\label{sec:users}
To complement the empirical results presented in the previous section, which demonstrate that our proposed experiment plan achieves marked efficiency benefits for complex experiments with overlapping pipelines,  we conducted a user study to evaluate how undergraduate students and postgraduate researchers experience experiment plans. The study was designed to assess users' understanding of the distinction between linear and tree‑based execution plans, evaluate the intuitiveness of the parameters introduced in \texttt{pt.Experiment}, examine the usefulness of the HTML visualisation in supporting comprehension of pipeline evaluation under the tree‑based experiment plan, and identify usability issues and areas for further improvement.

To this end, we designed
a short interactive Google Colab notebook in which participants first ran a baseline PyTerrier experiment using the default linear execution plan, both with and without prefix precomputation, and recorded the corresponding execution times. The final task required participants to determine for themselves how to enable the tree‑based plan within {\tt pt.Experiment}, specifically by identifying and adding the appropriate parameter. This task was designed to assess how intuitive the transition from linear to tree execution felt in practice. The notebook also asked participants to explore the visualisation features of the tree plan, allowing them to observe how pipelines branch and where computation is shared.

After completing these tasks, participants were asked to complete an anonymous questionnaire through Microsoft Forms. The questionnaire included a background question on their level of study and a set of Likert‑scale questions evaluating their familiarity with PyTerrier, the clarity of the transition from linear to tree‑based execution, and the usefulness of the visualisations. It also had free‑text questions asking participants to reflect on challenges encountered, CPU times, visualisation preferences, and potential improvements to parameter names, clarity, and usability.

\looseness -1 Our study involved a total of 11 participants -- 6 undergraduate and 5 postgraduate research students -- all of whom had some prior experience with IR. In particular, all of the undergraduate students were enrolled in the University of Glasgow's Information Retrieval course, which uses PyTerrier for running assessed coursework experiments involving sparse retrieval, pseudo-relevance feedback and learning-to-rank experiments.

From the questionnaire results, all users agreed or strongly agreed that they were familiar with the PyTerrier framework. This confirmed the study sample possessed sufficient expertise to meaningfully evaluate the proposed system. Across the task-based evaluation, users consistently reported that the notebook improved their conceptual understanding. The transition from linear to tree-based execution was rated as highly intuitive, with no participant reporting difficulty.  All responses were positive with comments like \emph{``None, it was pretty simple to change one parameter setting given the starter code."}

\looseness -1 The HTML visualisation received particularly positive feedback. 82\% of the participants strongly agreed that it helped them understand the behaviour of the tree-based experiment plan. One participant said: \emph{``I preferred the tree-based visualisation as I learn more from visual representations (I have terrible reading attention span)"}. Several participants emphasised that the visualisation made the optimisation's purpose immediately clear. Participants appreciated seeing how pipelines diverged and where computation was reused, noting that this clarified the rationale behind the optimisation. One participant even remarked that it \emph{“Looks cool!”}, suggesting that its aesthetics contribute positively to the overall user experience.

Execution time comparisons further reinforced the benefits of the tree-based plan. Participants using the larger MSMARCO dataset consistently observed reductions from approximately 6 minutes (with prefix precomputation) to around 4-5 minutes (tree‑based). Even those using smaller datasets reported the same trend. Every participant observed measurable improvements, demonstrating the robustness of the optimisation across different workloads.

Overall, participants found the study intuitive and well-structured. Many highlighted that \emph{``the structure of the tasks was clear and progressive"}, making it easy to follow the transition from linear to tree-based execution. While some users noted longer runtimes when working with the larger dataset, these issues were attributed to dataset size rather than the experiment plan itself. 

Overall, the responses demonstrate that the tree experiment achieves its objectives: all of the users understood the rationale behind the tree‑based execution plan, benefited from the visualisation, and observed clear improvements in execution efficiency.

\section{Conclusions}\label{sec:conc}
\looseness-1 Retrieval pipelines in IR are becoming increasingly complex, making it essential to support systematic experimentation with different pipeline instantiations for effective and efficient deployment. While PyTerrier offers a declarative language for formulating pipelines and evaluation, its current execution strategy can have marked efficiency disadvantages due to repeated execution of common retrieval stages. 

To address this limitation, this paper proposes considering the set of pipelines in an experiment as a DAG that must be traversed for execution -- a so-called {\em experiment plan}; going further we propose that a radix tree is an ideal data structure to easily formulate and traverse such a DAG. Experiments on MSMARCO v1 and v2 demonstrate the efficiency benefit of our radix tree, resulting in experiment duration reduction of upto 26\%. Furthermore, a small user study of undergraduate and research students indicated that the visualisation improved understanding, complementing the efficiency advantages. 

\enlargethispage{1em}Our demonstration experiment only addressed one possible pipeline configuration: retriever followed by pointwise and pairwise rankers but the approach generalises to a wide range of experimental configurations. These include (e.g.\ LLM-based) query reformulations, pseudo-relevance feedback, multi-stage generative retrieval and sequential RAG, which we have not shown in this paper but stand to benefit from our tree experiment plan.

\looseness -1 The trie-based experiment plan proposed in this work has already been deployed in PyTerrier. Currently PyTerrier detects pipelines with overlapping components, and suggests applying the trie-based experiment plan. However, as noted above, we may change the default configuration to be a trie-based experiment plan in the future.

For future work, we will look to further optimise experiment plans. One direction for exploration is to build on the {\em compilation} operations already supported by PyTerrier transformers~\cite{10.1145/3409256.3409829}, such as pushing rank‑cutoff operators ahead of rank‑invariant transformers or incorporating rank‑cutoffs directly into retrievers (such that {\tt BM25 \% 10} becomes {\tt BM25(k=10)}). We acknowledge that the interplay between pipeline compilation and experiment planning requires more investigation. For instance, two compiled pipelines respectively commencing with {\tt BM25(k=10)} and {\tt BM25(k=20)} no longer share a common prefix, and could not be combined using the trie-based techniques proposed in this paper. Addressing this limitation would require extending the transformer interface to detect when one transformer's output is a ranked superset of another's, enabling the smaller result to be derived from the larger one.

\begin{acknowledgments}
We acknowledge Sean MacAvaney for his helpful comments, and Martin Potthast for his initial suggestion of considering tree data structures during the presentation of~\cite{macavaney2025precomputation}. We also acknowledge the efforts of our user study participants, and the constructive comments of the reviewers.
\end{acknowledgments}

\section*{Declaration on Generative AI}
  
 During the preparation of this work, the authors used GPT and Grammarly for proof reading and spell checking. After using these tools, the authors reviewed and edited the content as needed and take full responsibility for the publication’s content. 

\bibliography{sample-ceur}

@String{Computing = "Computing" }

@String{Computer = "{IEEE} Computer" }

@String{Springer = "Springer-Verlag" }

@misc{R,
    title = {R: A Language and Environment for Statistical Computing},
    author = {{R Core Team}},
    organization = {R Foundation for Statistical Computing},
    address = {Vienna, Austria},
    year = {2019},
    url = {https://www.R-project.org/},
}

@inproceedings{macdonald2025constructing,
  title = {{Constructing and Evaluating Declarative RAG Pipelines in PyTerrier}},
  author = {Craig Macdonald and Jinyuan Fang and Andrew Parry and Zaiqiao Meng},
  year = {2025},
  publisher = {ACM},
  pages = {4035-4040},
  doi = {10.1145/3726302.3730150},
  booktitle = {Proceedings of the 48th International ACM SIGIR Conference on Research and Development in Information Retrieval}
}

@inproceedings{DBLP:conf/ecir/PenhaCH22,
  author       = {Gustavo Penha and
                  Arthur C{\^{a}}mara and
                  Claudia Hauff},
  
  title        = {Evaluating the Robustness of Retrieval Pipelines with Query Variation
                  Generators},
  booktitle    = {Advances in Information Retrieval - 44th European Conference on {IR}
                  Research, {ECIR} 2022, Stavanger, Norway, April 10-14, 2022, Proceedings,
                  Part {I}},
  series       = {Lecture Notes in Computer Science},
  pages        = {397--412},
  publisher    = {Springer},
  year         = {2022},
  doi          = {10.1007/978-3-030-99736-6\_27}
}

@ARTICLE{382293,
  author={Straube, D.D. and Ozsu, M.T.},
  journal={IEEE Transactions on Knowledge and Data Engineering}, 
  title={Query optimization and execution plan generation in object-oriented data management systems}, 
  year={1995},
  volume={7},
  number={2},
  pages={210-227},
  doi={10.1109/69.382293}}

@article{sellis1988multiple,
  title={Multiple-query optimization},
  author={Sellis, Timos K},
  journal={ACM Transactions on Database Systems (TODS)},
  volume={13},
  number={1},
  pages={23--52},
  year={1988},
  publisher={ACM New York, NY, USA}
}

@inproceedings{10.1145/3409256.3409829,
author = {Macdonald, Craig and Tonellotto, Nicola},
title = {Declarative Experimentation in Information Retrieval using PyTerrier},
year = {2020},
isbn = {9781450380676},
publisher = {Association for Computing Machinery},
address = {New York, NY, USA},
doi = {10.1145/3409256.3409829},
booktitle = {Proceedings of the 2020 ACM SIGIR on International Conference on Theory of Information Retrieval},
pages = {161–168},
numpages = {8},
location = {Virtual Event, Norway},
series = {ICTIR '20}
}

@inproceedings{macavaney2022adaptive,
author = {MacAvaney, Sean and Tonellotto, Nicola and Macdonald, Craig},
title = {Adaptive Re-Ranking with a Corpus Graph},
year = {2022},
isbn = {9781450392365},
publisher = {Association for Computing Machinery},
address = {New York, NY, USA},
doi = {10.1145/3511808.3557231},
booktitle = {Proceedings of the 31st ACM International Conference on Information \& Knowledge Management},
pages = {1491–1500},
numpages = {10},
location = {Atlanta, GA, USA},
series = {CIKM '22}
}

@inproceedings{tonellotto2013efficient,
  title = {Efficient and effective retrieval using selective pruning},
  author = {Nicola Tonellotto and Craig Macdonald and Iadh Ounis},
  year = {2013},
  publisher = {ACM},
  pages = {63-72},
  doi = {10.1145/2433396.2433407},
  booktitle = {Proceedings of the sixth ACM international conference on Web search and data mining}
}

@misc{pradeep2023rankvicunazeroshotlistwisedocument,
      title={{RankVicuna}: Zero-Shot Listwise Document Reranking with Open-Source Large Language Models}, 
      author={Ronak Pradeep and Sahel Sharifymoghaddam and Jimmy Lin},
      year={2023},
      eprint={2309.15088},
      archivePrefix={arXiv},
      primaryClass={cs.IR},
      url={https://arxiv.org/abs/2309.15088}, 
}

@article{DBLP:journals/corr/abs-2101-05667,
  author       = {Ronak Pradeep and
                  Rodrigo Nogueira and
                  Jimmy Lin},
  title        = {The Expando-Mono-Duo Design Pattern for Text Ranking with Pretrained
                  Sequence-to-Sequence Models},
  journal      = {CoRR},
  volume       = {abs/2101.05667},
  year         = {2021},
  url          = {https://arxiv.org/abs/2101.05667},
  eprinttype   = {arXiv},
  eprint       = {2101.05667}
}

@inproceedings{macdonald2026replicability,
  title     = {A Replicability Study of Joint Product Quantisation for Effective Space-Efficient Dense Retrieval},
  author    = {Macdonald, Craig and Shen, Zhili and Tonellotto, Nicola},
  booktitle = {Proceedings of the 49th International ACM SIGIR Conference on Research and Development in Information Retrieval},
  year      = {2026},
 doi = {10.1145/3805712.3808565},
 note = {In press}
}

@inproceedings{irdatasets,
author = {MacAvaney, Sean and Yates, Andrew and Feldman, Sergey and Downey, Doug and Cohan, Arman and Goharian, Nazli},
title = {Simplified Data Wrangling with ir\_datasets},
year = {2021},
isbn = {9781450380379},
publisher = {Association for Computing Machinery},
address = {New York, NY, USA},
doi = {10.1145/3404835.3463254},

booktitle = {Proceedings of the 44th International ACM SIGIR Conference on Research and Development in Information Retrieval},
pages = {2429–2436},
numpages = {8},
location = {Virtual Event, Canada},
series = {SIGIR '21}
}

@inproceedings{10.1145/3459637.3482013,
author = {Macdonald, Craig and Tonellotto, Nicola and MacAvaney, Sean and Ounis, Iadh},
title = {{PyTerrier}: Declarative Experimentation in {Python} from {BM25} to Dense Retrieval},
year = {2021},
isbn = {9781450384469},
publisher = {Association for Computing Machinery},
address = {New York, NY, USA},
doi = {10.1145/3459637.3482013},
booktitle = {Proceedings of the 30th ACM International Conference on Information \& Knowledge Management},
pages = {4526–4533},
numpages = {8},
location = {Virtual Event, Queensland, Australia},
series = {CIKM '21}
}

@inproceedings{DBLP:conf/ecir/LionisMM26,
  author       = {Emmanouil Georgios Lionis and
                  Craig Macdonald and
                  Sean MacAvaney},
  
  title        = {Pipeline Inspection, Visualization, and Interoperability in {PyTerrier}},
  booktitle    = {Advances in Information Retrieval - 48th European Conference on Information
                  Retrieval, {ECIR} 2026, Delft, The Netherlands, March 29 - April 2,
                  2026, Proceedings, Part {IV}},
  series       = {Lecture Notes in Computer Science},
  pages        = {159--165},
  publisher    = {Springer},
  year         = {2026},
  doi          = {10.1007/978-3-032-21321-1\_23}
}

@article{craswell2019overview,
    author = "Craswell, Nick and Mitra, Bhaskar and Yilmaz, Emine and Campos, Daniel and Voorhees, Ellen M.",
    title = "Overview of the {TREC} 2019 deep learning track",
    journal = "CoRR",
    volume = "abs/2003.07820",
    year = "2020",
    url = "https://arxiv.org/abs/2003.07820"
}

@inproceedings{wang2011cascading,
author = {Wang, Lidan and Lin, Jimmy and Metzler, Donald},
title = {A cascade ranking model for efficient ranked retrieval},
year = {2011},
isbn = {9781450307574},
publisher = {Association for Computing Machinery},
address = {New York, NY, USA},
doi = {10.1145/2009916.2009934},
booktitle = {Proceedings of the 34th International ACM SIGIR Conference on Research and Development in Information Retrieval},
pages = {105–114},
numpages = {10},
location = {Beijing, China},
series = {SIGIR '11}
}

@article{fntir2018efficient,
    author = {Tonellotto, Nicola and Macdonald, Craig and Ounis, Iadh},
    title = {Efficient Query Processing for Scalable Web Search},
    journal = {Foundations and Trends in Information Retrieval},
    volume = {12},
    number = {4-5},
    pages = {319-500},
    year = {2018},
    month = {11},
    issn = {1554-0669},
    doi = {10.1561/1500000057}
}

@article{DBLP:journals/ftir/Liu09,
  author       = {Tie{-}Yan Liu},
  title        = {Learning to Rank for Information Retrieval},
  journal      = {Found. Trends Inf. Retr.},
  volume       = {3},
  number       = {3},
  pages        = {225--331},
  year         = {2009},
  doi          = {10.1561/1500000016},

}

@inproceedings{macavaney2025precomputation,
  author       = {Sean MacAvaney and
                  Craig Macdonald},
  title        = {On Precomputation and Caching in Information Retrieval Experiments
                  with Pipeline Architectures},
  booktitle    = {Proceedings of the 2nd International Workshop on Open Web Search co-located
                  with the 47th European Conference on Information Retrieval {(ECIR}
                  2025), Lucca, Italy, April 10, 2025},
  series       = {{CEUR} Workshop Proceedings},
  pages        = {23--35},
  publisher    = {CEUR-WS.org},
  year         = {2025},
  url          = {https://ceur-ws.org/Vol-4137/WOWS\_2025\_paper\_3.pdf}
}

@inproceedings{harizopoulos2005qpipe,
author = {Harizopoulos, Stavros and Shkapenyuk, Vladislav and Ailamaki, Anastassia},
title = {QPipe: a simultaneously pipelined relational query engine},
year = {2005},
isbn = {1595930604},
publisher = {Association for Computing Machinery},
address = {New York, NY, USA},
doi = {10.1145/1066157.1066201},
booktitle = {Proceedings of the 2005 ACM SIGMOD International Conference on Management of Data},
pages = {383–394},
numpages = {12},
location = {Baltimore, Maryland},
series = {SIGMOD '05}
}

@ARTICLE{9894699,
  author={Aseman-Manzar, Mohammad-Mohsen and Karimian-Aliabadi, Soroush and Entezari-Maleki, Reza and Egger, Bernhard and Movaghar, Ali},
  journal={IEEE Transactions on Services Computing}, 
  title={Cost-Aware Resource Recommendation for {DAG}-Based Big Data Workflows: An {Apache Spark} Case Study}, 
  year={2023},
  volume={16},
  number={3},
  pages={1726-1737},
  doi={10.1109/TSC.2022.3203010}}

@phdthesis{Neumann2005a,
  author       = {Thomas Neumann},
  title        = {Efficient generation and execution of DAG-structured query graphs},
  school       = {University of Mannheim, Germany},
  year         = {2005},
  url          = {https://api.semanticscholar.org/CorpusID:487710}
}

@inproceedings{FrobeRMDRB0HP23,
  author       = {Maik Fr{\"{o}}be and
                  Jan Heinrich Reimer and
                  Sean MacAvaney and
                  Niklas Deckers and
                  Simon Reich and
                  Janek Bevendorff and
                  Benno Stein and
                  Matthias Hagen and
                  Martin Potthast},

  title        = {The Information Retrieval Experiment Platform},
  booktitle    = {Proceedings of the 46th International {ACM} {SIGIR} Conference on
                  Research and Development in Information Retrieval, {SIGIR} 2023, Taipei,
                  Taiwan, July 23-27, 2023},
  pages        = {2826--2836},
  publisher    = {{ACM}},
  year         = {2023},
  doi          = {10.1145/3539618.3591888}
}

@inproceedings{sparkterrier,
author = {Macdonald, Craig},
title = {Combining {Terrier} with {Apache} {Spark} to create Agile Experimental Information Retrieval Pipelines},
year = {2018},
isbn = {9781450356572},
publisher = {Association for Computing Machinery},
address = {New York, NY, USA},
doi = {10.1145/3209978.3210174},
booktitle = {The 41st International ACM SIGIR Conference on Research \& Development in Information Retrieval},
pages = {1309–1312},
numpages = {4},
location = {Ann Arbor, MI, USA},
series = {SIGIR '18}
}

@inproceedings{sparksql,
author = {Armbrust, Michael and Xin, Reynold S. and Lian, Cheng and Huai, Yin and Liu, Davies and Bradley, Joseph K. and Meng, Xiangrui and Kaftan, Tomer and Franklin, Michael J. and Ghodsi, Ali and Zaharia, Matei},
title = {Spark {SQL}: Relational Data Processing in {Spark}},
year = {2015},
isbn = {9781450327589},
publisher = {Association for Computing Machinery},
address = {New York, NY, USA},
doi = {10.1145/2723372.2742797},
booktitle = {Proceedings of the 2015 ACM SIGMOD International Conference on Management of Data},
pages = {1383–1394},
numpages = {12},
location = {Melbourne, Victoria, Australia},
series = {SIGMOD '15}
}

@Inbook{Karimov2020,
author="Karimov, Elshad",
title="Trie Data Structure",
bookTitle="Data Structures and Algorithms in Swift: Implement Stacks, Queues, Dictionaries, and Lists in Your Apps",
year="2020",
publisher="Apress",
address="Berkeley, CA",
pages="67--75",
isbn="978-1-4842-5769-2",
doi="10.1007/978-1-4842-5769-2_9"
}

@inproceedings{gunasinghe2013adaptive,
  title={The adaptive suffix tree: A space efficient sequence learning algorithm},
  author={Gunasinghe, Upuli and Alahakoon, Damminda},
  booktitle={The 2013 International Joint Conference on Neural Networks (IJCNN)},
  pages={1--8},
  year={2013},
  organization={IEEE},
  doi={10.1109/IJCNN.2013.6707052}
}

@article{Morrison68,
  author       = {Donald R. Morrison},
  title        = {{PATRICIA} - Practical Algorithm To Retrieve Information Coded in
                  Alphanumeric},
  journal      = {J. {ACM}},
  volume       = {15},
  number       = {4},
  pages        = {514--534},
  year         = {1968},
  doi          = {10.1145/321479.321481}
}

@book{larsson1999structures,
  title={Structures of String Matching and Data Compression.},
  author={Larsson, N Jesper},
  year={1999},
  publisher={Lund University, Sweden}
}

@article{nguyen2016ms,
author = {Nguyen, Tri and Rosenberg, Mir and Song, Xia and Gao, Jianfeng and Tiwary, Saurabh and Majumder, Rangan and Deng, Li},
title = {{MS MARCO: A Human Generated MAchine Reading COmprehension Dataset}},
year = {2016},
month = {November},
url = {https://www.microsoft.com/en-us/research/publication/ms-marco-human-generated-machine-reading-comprehension-dataset/},
}

@misc{craswell2025overviewtrec2021deep,
      title={Overview of the {TREC} 2021 deep learning track}, 
      author={Nick Craswell and Bhaskar Mitra and Emine Yilmaz and Daniel Campos and Jimmy Lin},
      year={2025},
      archivePrefix={arXiv},
      primaryClass={cs.IR},
      url={https://arxiv.org/abs/2507.08191}, 
}

\end{document}